\documentstyle[12pt,dpf]{article}
\title{\bf A NEW ACTION FOR HEAVY LATTICE FERMIONS\thanks{Talk given at
the 7th Meeting of the APS Division of Particles and Fields,
Fermilab, Nov. 10-14, 1992.
}
}
\author{PAUL B. MACKENZIE \\[0.3cm]
{\it Theoretical Physics Group, Fermi National Accelerator  Laboratory,} \\
{\it P.O. Box 500, Batavia, IL 60510, USA}}
\abstract{
I describe a unified formalism for lattice fermions,
in which the relativistic action of Wilson and the
 nonrelativistic  and static actions
appear as special cases.
It is valid at all values of $m_q a$, including $m_q  a \approx 1$.
In the limit $m_q a \ll 1 $, the formulation reduces to the light
quark action of Wilson.
In the limit $m_q a \gg 1 $, the formulation reduces to
the nonrelativistic action of  Thacker and Lepage, and to the static
action of Eichten.
  }
\begin{document}
%%\vfill
\maketitle
%-----------------------------------------------------------------------
%%\vfill \newpage
%-----------------------------------------------------------------------

%\subsection{A New Action for Four Component Fermions}\label{general}
Present and future
lattice calculations involving $b$ and $c$ quarks include some of the
most important applications of lattice gauge theory to standard model
physics.
These include the heavy meson decay constants, the $B \overline{B}$
mixing amplitude, and various semileptonic decay amplitudes,
which are all crucial in extracting CKM angles from experimental data.
They also include the extraction of $\alpha_s$ from the charmonium
and bottomonium spectra.

There exist two main classes of methods for lattice fermions:
small mass expansions and large mass expansions.
  The standard
Wilson action contains errors which vanish as powers of $m_q a$,
the quark mass in lattice units.  This series of corrections does not
converge when $m_q a > 1.$
There  exists a second class of methods for treating lattice fermions when
$m_q a > 1$: Nonrelativistic QCD (NRQCD)\cite{NRQCD} and the static
approximation.\cite{static}
Because coefficients of higher terms in the  Lagrangians of these methods
 (such as
${\bf D}^2/{(2 m)}$) are explicit functions of $1/m$, the loop
corrections are also explicit functions of
$1/m$.  These begin to diverge as $m a$ is reduced below a value of
order one, making the nonrelativistic expansion impractical.
The masses of the $b$ and $c$ quarks are such that $m_q a$ is $O(1)$ at
the lattice spacings used in current numerical work.
Therefore,  calculations of such crucially interesting
quantities as the heavy meson decay constants $f_B$ and $f_D$ have
often involved awkward interpolations between results in the static
approximation and results using Wilson fermions through a region where
neither approximation is well behaved.
 While such an
approach is probably workable, it is clearly desirable to have a
method for lattice fermions which is well behaved throughout the
region of interest.

Since the large mass and small mass formalisms are both descriptions
of QCD, it is not surprising that they share certain fundamental features.
  Like Wilson fermions ($\psi$), the fermions of
NRQCD contain four components per site: a two-component quark field
($\phi$) and a two-component antiquark field ($\chi$).
Further, the two actions employ the same sorts of interactions:
covariant time derivatives, covariant Laplacians, etc.
To find a formalism uniting the two actions,
 we therefore consider the following  generalized
Lagrangian:\cite{KronMack,lat92}
\begin{eqnarray}
{\cal L}&=& \phi^* (\ \ \ c_1 \ \Delta_t^- + m_0 - \frac{c_2}{2} \sum_{i}
\Delta^+_i\Delta^-_i)\phi_n \nonumber + c_3 \ \phi^*
\sum_{i} \sigma_i \Delta_i \chi_n \\ &+& \chi^* (- c_1 \ \Delta_t^+ +
m_0 - \frac{c_2}{2} \sum_{i} \Delta^+_i\Delta^-_i)\chi_n +
 c_3 \ \chi^* \sum_{i} \sigma_i \Delta_i \phi_n.
\end{eqnarray}
($\Delta^+$, $\Delta^-$, and $\Delta$ are the forward, backward, and
symmetric discrete difference operators, respectively.
I will use $m_q$ for the physical quark mass and $m_0$ for the bare
quark mass on the lattice.)
With the choice of parameters $c_1=c_2=c_3=1,$ this is simply
 the standard Wilson action.
When $c_1=1$ (times a correction factor when $m a \gg 1$),
$c_2=\frac{1}{m}$, and $c_3$ is negligible, it is a good, if somewhat
unconventional, Lagrangian for NRQCD.
The  bare mass is conventionally
omitted in NRQCD calculations, but we are free to leave it in the
theory.  The usual Dirac coupling between quarks and
antiquarks is absent (having been transformed into higher derivative
interactions by the Foldy-Wouthuysen transformation), but we may add
back a sufficiently suppressed amount of this interaction without
spoiling the nonrelativistic theory.
It is thus possible to
adjust the parameters for this particular NRQCD action
in such a way that as $m_0$ is reduced, instead
of blowing up, the theory turns smoothly into the Wilson theory.
One must find appropriate normalization conditions to determine the
parameters of the action which lead to the Wilson theory for
$a m_q \ll 1$, and to NRQCD and the static approximation for
$a m_q \ll 1$, and which work at all values of $a m_q$.

It is illuminating in this regard to expand the equation for Wilson propagators
nonrelativistically when the mass is large,
to see what breaks down.
 After normalizing the
fields by $\frac{1}{\sqrt{1-6\kappa}}$\cite{lunorm} (not
$\frac{1}{\sqrt{2\kappa}}$ as is commonly used) one may obtain
\begin{equation}\label{SE}
0 = \left[{\cal -E + M}  + (1-U^\dagger_{{\bf n},0} )
 -\frac{1}{2}\left( \frac{1}{ m_0 } +
\frac{1}{(1+m_0)\ (2+m_0) } \right)  \sum_{i} ( \Delta_i )^2
\right]\phi_{\bf n} ,
\end{equation}
where $\cal E$ is the energy eigenvalue obtained from the transfer
matrix and ${\cal M=E}_{p^2=0} = \ln(1+m_0)$.  This is a lattice
Schr\"odinger equation not unlike the one obtained from NRQCD, but it
has some unusual features.  Most important, the two ``masses'' in the
equation, the rest mass ${\cal M}= \ln(1+m_0)$,
 and the mass governing  the energy-momentum relation
$\frac{1}{M} = \frac{1}{ m_0 } +
\frac{1}{(1+m_0)\ (2+m_0) }$, are completely different.
$\cal M$ plays little dynamical role in heavy quark systems and is
usually omitted from NRQCD and static approximation
calculations.  The dynamically more important
condition $\partial {\cal E}/ \partial
p^2 = {1}/{(2m_q)}$ is used to fix the mass in NRQCD.
This condition is also the same as the usual mass condition
for Wilson fermions when $a m \ll 1$,
since when $m_0 a $ is small,
\begin{equation}
 \frac{1}{ m_0 } +
\frac{1}{(1+m_0)\ (2+m_0) } \approx \frac{1}{\cal M} \approx \frac{1}{m_0}.
\end{equation}
 However,
if the rest mass $\cal M$ is used to fix the fermion mass for Wilson fermions
when $a m > 1$, the energy-momentum
mass condition $\partial {\cal E}/ \partial
p^2 = {1}/{(2m_q)}$, which is usually more important,
 will be completely incorrect.
The two masses can be put back into agreement with the use of the
Lagrangian\cite{KronMack,lat92}
\begin{eqnarray} \label{newaction}
{\cal L} =  -\bar{\psi}_{n} \psi_{n}  &+&\kappa_t
\bar{\psi}_{n}(1-\gamma_0) U_{n,0}
\psi_{n+\hat{0}} + h. c.		\nonumber\\
&+&\kappa_s\sum_{i}\bar{\psi}_{n}(1-\gamma_i) U_{n,i}
\psi_{n+\hat{i}} + h. c.
\end{eqnarray}

Thus, it seems that an action closely related to the Wilson action is
a member of the class of actions suitable for NRQCD.
Further,  in NRQCD and in the static approximation, $\cal M$ plays little
dynamical role. It can usually be ignored, and is usually omitted.
This suggests that for that majority of calculations in which the
mass gap between states containing different numbers of quarks is
unimportant,
the standard Wilson action itself can be used when
$a m > 1$ as long as $\cal M$ is ignored and $\partial {\cal E}/
\partial p^2 = \frac{1}{2m}$ is used to fix the quark mass, as is done
in NRQCD.

This proposal is obviously correct in free field theory, where we can
calculate the behavior of quark propagators exactly to see that the
proposed interpretation makes sense.  It is certainly correct in mean
field theory, too.  Mean field improvement of these fermions, as of
Wilson fermions, is simply the absorption of a ``mean link'' $u_0$
 into an effective $\tilde{\kappa} \equiv u_0
\kappa$ and then proceeding as with free field theory.\cite{PT}
  (A plausible
estimate of the mean link in this context is probably $u_0 \approx
1/8\kappa_c$.)  It remains to be shown whether the theory is somehow
spoiled by renormalization.

Perturbatively, Green functions must be expanded in $p^2$ and
$\alpha_s$.  The coefficient of
each term in the expansion is an explicit function of the
quark mass, since the theory must be solved exactly in $m a$. (The is
also the case for the loop corrections of NRQCD.\cite{BAT}) If these
functions become singular or badly behaved in some way, the theory
could conceivably break down.  The one loop perturbative corrections
contain all of the ugliest features of Wilson and NRQCD perturbation
theory simultaneously, and have only been begun.  There is, however,
one numerical calculation by El-Khadra\cite{cnorm} indicating that
nothing too surprising occurs.  The one-loop correction to the local
current normalization for Wilson fermions with the naive normalization
is\cite{Zhang}
\begin{equation}   \label{currentnorm1}
\langle \psi | V_4^{loc} | \psi \rangle =
 \frac{1}{2 \kappa (1-0.17 g^2) }.
\end{equation}
The correct normalization with mean field improvement is
\begin{equation}   \label{currentnorm}
\langle \psi | V_4^{loc} | \psi \rangle =
 \frac{1}{(1-\frac{6\kappa}{8\kappa_c})(1-0.06g^2) }.
\end{equation}
The remaining perturbative correction, $0.06g^2$, becomes an explicit
(so far uncalculated) function of $m$ (or $\kappa$) in the new
formalism.  The small mass limit of this function is $0.06g^2$.  This function
 must not become singular if the theory is to make
sense.
This normalization is straightforward to calculate numerically.
 As described in Ref. \cite{cnorm},
the nonperturbative calculation agrees with Eq. \ref{currentnorm} to within
a few per cent.
It disagrees with Eq. \ref{currentnorm1} by around a factor of two.
It can be seen that for this
quantity, not only is the unknown function of $m$ not singular, it is
approximately equal to 1.

Putting the new action on a secure footing will ultimately require: 1)
determination of the bare parameters of the action with mean field
theory and full perturbation theory, 2) nonperturbative tests of the
perturbative results, and 3) phenomenological tests of the resulting
action in calculations of well understood physical quantities.  Not
much of this program has yet been accomplished.
However, from what is known now, there appears to be no insuperable
obstacle  to developing a formulation of lattice fermions practical for
all values of the quark mass, not just $m_q a \gg 1$ or
$m_q a \ll 1.$

Care will clearly be required in formulating normalization conditions
which capture the most important physics in both the relativistic and
nonrelativistic regions.  (Identifying $\partial {\cal E}/ \partial
p^2$ rather than $\cal M$ as the fundamental mass condition is example
number one of these.)

\vspace{1.0em} % \section*{Acknowledgements}
I thank G.P. Lepage and my collaborators A. X. El-Khadra and A. S. Kronfeld
 for useful discussions.
 Fermilab is operated by Universities Research Association Inc.\ under %%
 contract with the U.S. Department of Energy.%%

This material appeared in slightly altered form in Ref. \cite{lat92}.

\end{document}